\documentclass[aps,pra,groupedaddress,reprint]{revtex4-1}
\usepackage[utf8x]{inputenc}
\usepackage{bbm} 
\usepackage{amsfonts,amsmath,amssymb,stmaryrd}

\usepackage{graphicx}
\usepackage{subfigure}  
\usepackage{bbm} 
\usepackage{hyperref}

\usepackage{verbatim}

\newcommand{\Ham}{\mathcal{H}}

\DeclareMathSymbol{\epsilon}{\mathord}{letters}{"22}
\DeclareMathSymbol{\theta}{\mathord}{letters}{"23}
\DeclareMathSymbol{\rho}{\mathord}{letters}{"25}
\DeclareMathSymbol{\phi}{\mathord}{letters}{"27}

\DeclareMathSymbol{\varepsilon}{\mathord}{letters}{"0F}
\DeclareMathSymbol{\vartheta}{\mathord}{letters}{"12}
\DeclareMathSymbol{\varphi}{\mathord}{letters}{"1E}
\DeclareMathSymbol{\varrho}{\mathord}{letters}{"1A}

\DeclareMathOperator{\Tr}{Tr}

\begin{document}

\title{Critical exponents of steady-state phase transitions in fermionic lattice models}

\author{M. H\"oning}
\author{M. Moos}
\author{M. Fleischhauer}

\affiliation{Department of Physics and research center OPTIMAS, University of Kaiserslautern, Germany}

\date{\today}

\begin{abstract}
We discuss reservoir induced phase transitions of lattice fermions in the non-equilibrium steady state
(NESS) of an open system with local reservoirs. These systems may become critical in the sense
of a diverging correlation length upon changing the reservoir coupling. We here show that the
transition to a critical state is associated with a vanishing gap in the damping spectrum. 
It is shown that although in linear systems there can be a transition to a critical state there
is no reservoir-induced quantum phase transition between distinct phases with non-vanishing
damping gap.  
We derive the static and dynamical critical exponents corresponding to the transition to a
critical state
and show that their possible values, defining universality classes of reservoir-induced phase transitions 
are determined by the coupling range of the independent local reservoirs. If a reservoir couples to $N$ neighboring
lattice sites, the critical exponent can assume all fractions from 1 to $1/(N-1)$. 
\end{abstract}

\maketitle

\section{Introduction}

Experiments with cold atoms allow unmatched control over quantum systems giving access to a number of interesting many-body effects \cite{Bloch2008}. A prominent example are quantum phase transitions \cite{Sachdev1999} in the ground state of many-body Hamiltonians. On the other hand the tools of quantum optics allow to control the interaction with 
the environment and systems can be prepared in the non-equilibrium steady state (NESS) of an open dynamics instead \cite{Diehl2008, Verstraete2009, Mueller2012,PhysRevA.84.043637}.
In general the steady state of an open system is very different from the ground state of the corresponding Hamiltonian or even from thermal states and thus driving a system into that state is uniquely different from the canonical treatment of decoherence as a perturbing effect on ground states \cite{Pichler2010}. As the steady state or more generally the stationary  space is an attractor of the nonunitary evolution, it is robust against further decoherence. Such a scheme, where the stationary state is a projector, i.e. a dark state of the nonunitary evolution, has recently been proposed e.g. by Diehl et. al. to prepare exotic states such as the Kitaev edge modes \cite{Diehl2011} or BCS-type states \cite{Diehl2010a}. The time evolution within a higher dimensional dark-space can correspond to interesting effective many-body Hamiltonians. Examples include the recently observed dissipative Tonks-Girardeau gas of atoms \cite{Duerr2009} or corresponding proposals for photons \cite{Kiffner2010}.

As there is a growing theoretical and experimental interest in engineered open systems, see e.g.
the recent review \cite{Mueller2012}, we want to discuss in the present paper the analogue to a quantum phase 
transition in Hamiltonian systems in the NESS of an open system, described by a Lindblad Liouville operator 
and induced by changing reservoir couplings. In Hamiltonian systems a quantum phase transition results
from the competition of two non-commuting parts of a microscopic Hamiltonian $H=H_1+g H_2$ with different
symmetries upon changing their relative strength $g$ \cite{Sachdev1999}. A transition
between two distinct quantum phases occurs at a critical value $g=g_c$ and can be identified by a non-analytic
behavior of an order parameter. Furthermore at the critical point the excitation gap $\Delta_H$, i.e. the
energy difference between the lowest excited state of $H$ and the ground state, closes.
At the same time certain correlations become infinitely long-ranged indicated by a diverging correlation
length $\xi$. Criticality induced by reservoirs in open many-body systems has recently been
discussed by Eisert and Prosen \cite{Eisert2010b} for free systems, where criticality was defined in
terms of a diverging correlation length. Here we will reexamine free fermionic lattice models as generic models to study noise-induced quantum phase transitions. As analogue to the excitation gap $\Delta_H$ in
unitary systems we will consider the damping gap $\Delta$ of the Liouvillian as indicator of a critical point for reservoir-induced phase transitions. We will show that if the manifold of reservoir parameters 
is of dimension $d$ the values for which the damping gap closes and the system becomes critical 
is generically of dimension $d-2$ for free systems. As a consequence  
the gapped phases are always connected in parameter space and there are only 
isolated critical points or manifolds for reservoir-induced quantum phase transitions of linear
systems. 

An important aspect of quantum phase transitions in unitary systems is that 
close to criticality, microscopic details of the interaction become irrelevant leading to universal
behavior and allowing to classify phase transitions 
according to  universality classes. These classes are defined by common
critical exponents $\lambda$ and dynamical critical exponents $z$, characterizing the divergence
of the correlation length and the closure of the excitation gap when approaching the critical point:
\begin{eqnarray*}
\Delta_H \, &\sim& \, |g-g_c|^{z\lambda},\\
\xi^{-1} \, &\sim&\, |g-g_c|^{\lambda}.
\end{eqnarray*}
We here argue that similar universal exponents can be derived for reservoir-induced transitions
where the possible critical exponents depend solely on the coupling range of the independent local
reservoirs.

After introducing linear fermionic lattice models coupled to local reservoirs and
summarizing the general methods to derive the NESS of these systems in section II, we
discuss a specific realization of such a model with reservoirs coupling to two neighboring
lattice sites in cold atom systems in section III. 
Using this example as illustration we then formulate the key questions addressed in the remainder
of this  paper: Under what conditions do reservoir induced phase transitions occur and to what
universality classes can they be attributed to?
We will then show in section IV for general reservoir 
couplings that noise-induced criticality defined by a diverging
correlation length occurs at points in parameter space where the damping gap, i.e. the spectral gap
of relaxation rates of the Liouvillian dynamics closes. In contrast to Hamiltonian systems, where the
parameters that drive a phase transition are real, in Liouvillian system these parameters
are complex. We then show that for linear models criticality occurs only at isolated 
singularities or parameter manifolds of dimension $d-2$, where $d$ is the dimension of the reservoir
parameters. An immediate consequence of this is that there is only one connected phase with a
non-vanishing damping gap. We derive the possible
critical exponents, defining universality classes of noise-induced phase transitions and show that
they are simple fractions determined by the number of sites that couple to the same reservoir.
Finally  we establish a simple relation between the critical and dynamical critical exponents.

\section{free lattice fermions with linear reservoir coupling}

As a generic system  we study a free and translationally invariant fermionic chain described by the system Hamiltonian ${\cal H}$. In addition to the unitary evolution
there is a non-unitary interaction with reservoirs which is assumed to be Markovian. This
allows for an effective description of the dynamical equation for  the density operator $\rho$ via Lindblad operators $L_\mu$
\begin{align}
\begin{split}
\frac{d}{dt}\rho&=\mathcal{L}(\rho)\\
&=-i[\Ham,\rho]+\frac{1}{2}\sum_\mu\left(2L_{\mu}\rho L_{\mu}^\dagger-\{L_\mu^\dagger L_{\mu},\rho\}\right). \label{eqn:lindblad_master}
\end{split}
\end{align}
In the following we will use Majorana operators
\begin{equation}
 w_{2j-1}=\hat c_j^\dagger +\hat c_j,\qquad w_{2 j}= i(\hat c_j-\hat c_j^\dagger),
\end{equation}
instead of the more familiar fermionic creation and annihilation operators $\hat c_j^\dagger$ and $\hat c_j$, as the Majorana operators allow for an easy representation of the steady state of the considered linear systems. They are analogues of the bosonic position and momentum operators, are hermitian and 
fulfill a simple anti-commutation relation $\{w_j,w_k\}=2\delta_{j,k}$.
The signature of a free system is the bilinearity of the Hamiltonian 
and the linearity of Lindblad generators when expressed in terms of Majorana operators
\begin{eqnarray}
\Ham_S &=& \sum_{j,k=1}^{2L}(H)_{j,k}w_j w_k,\\
L_{\mu} &=&\sum_{j=1}^{L}(l)_{\mu,j,1}w_{2j-1}+(l)_{\mu,j,2}w_{2j}.
\end{eqnarray}
In the following we consider local reservoirs only. I.e. we restrict the reservoir coupling to a finite number of $N$ adjacent sites.
To explicitly take into account translation invariance we reformulate the Hamiltonian and the Lindblad generators
\begin{eqnarray}
 \Ham_S&=& \sum_{j}\tau_j\left(\sum_{m,n}(h)_{m,n}w_mw_n\right),\\
L_j&=&\tau_j\Bigl(\sum_{m=0}^{N-1} \nu_m e^{ig_m}w_{2m-1}\label{eqn:generators}\\
&&\qquad+\tilde\nu_{m} e^{i\tilde g_{m}}w_{2m}\Bigr),\quad j\in \mathbb{Z}\nonumber,
\end{eqnarray}
where we have introduced the operator $\tau_j$, which shifts a local operator by $j$ lattice sites.
One recognizes that the Lindblad generators contain $2 N$ independent complex
parameters $s_m = \nu_m e^{i g_m}$ and $q_m = \tilde\nu_{m} e^{i \tilde g_{m}}$. Only
$2N-1$ are relevant however in the context of reservoir induced phase transitions as one 
coefficient can be pulled out and only determines the overall time scale of the damping. 
Thus we may set without loss of generality $s_0=1$.

The dynamical equation (\ref{eqn:lindblad_master}) contains only quadratic terms in fermionic operators, thus 
the NESS $\rho_0$ is Gaussian \cite{Prosen2008} and is fully
described by the correlation matrix 
\begin{eqnarray}
 (\Gamma)_{j,k}=\frac{i}{2}\Tr\bigl\{\rho_0(w_j w_k-w_kw_j)\Bigr\}.
\end{eqnarray}
Higher order correlation functions can be calculated using Wick's theorem \cite{bravyi2005}. 
$\Gamma_{jk}$ is antisymmetric, has purely imaginary entries and is directly related to the Grassman representation of the steady state $w(\rho_0,\vartheta)=\frac{1}{2^L}\text{exp}\Big(-\frac{i}{2}\vartheta_j\Gamma_{jk}\vartheta_k\Big) $.
The dynamics generated by Eq.(\ref{eqn:lindblad_master}) take the simple form of a 
linear matrix differential 
equation for the correlation matrix $\Gamma$ \cite{Eisert2010b}
\begin{equation}
\frac{d}{dt}\Gamma=X^T \Gamma+\Gamma X-Y. \label{eqn:lindblad_cov}
\end{equation}
The matrices $X$ and $Y$ represent the Hamiltonian and the Lindblad generators $X=-4iH-(R+R^*)$ and $Y=2i(R-R^*)$ with the reservoir matrix $R=\sum_\mu l_{\mu}\otimes l_{\mu}^*$.

We are only interested in the steady state of the system given by the solution of the Lyapunov Sylvester equation
\begin{equation}
 X^T\Gamma_0+\Gamma_0 X=Y. \label{eq:Lyapunov} 
\end{equation}
For translational invariant and infinitely large systems, the correlation matrix of the steady state is circulant and can be represented by its Fourier
transform, called the symbol function $\gamma(\phi)$, which is a $2\times 2$ matrix corresponding to the two different types of Majorana species with even or odd indices:
\begin{equation}
 \gamma(\phi)=\left(\begin{array}{cc}
                     \gamma_{11}(\phi) & \gamma_{12}(\phi) \\
                     \gamma_{21}(\phi) & \gamma_{22}(\phi)
                    \end{array}\right)
\end{equation}
where
\begin{equation}
 \langle w_{2j-1} w_{2(j+d)-1}\rangle = \langle w_1 w_{1+2d}\rangle = \frac{1}{2\pi}
\int_0^{2\pi}\!\!d\phi  \,\, e^{i\phi d} \gamma_{11}(\phi)
\label{eq:correlation}
\end{equation}
and analogously for other types of Majorana operators.

In terms of the $2\times 2$ symbol functions the Lyapunov Sylvester equation reads
\begin{equation}
x(-\phi)^T\gamma(\phi)+\gamma(\phi)x(\phi)=y(\phi). \label{eqn:lindblad_steadyI}
\end{equation}
The matrices $x(\phi)$ and $y(\phi)$ are calculated in correspondence to the $X$ and $Y$ matrices of the finite size model. 

\section{a quantum-optical example: lattice fermions with two-site reservoir coupling}

\begin{figure}
\begin{center}
\includegraphics[width=0.90\columnwidth]{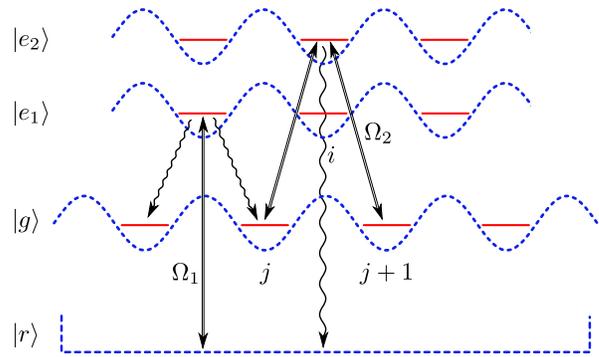}
\caption{(Color online) Fermions with four internal states are trapped in state dependent optical lattices in the above configuration to realize the reservoir coupling of two sites in the $|g\rangle$ lattice. The shallow potential for atoms in internal state $|r\rangle$ acts as the reservoir, which is coupled in a controlled way to $|g\rangle$. The effective pumping is driven by a laser on the $|e_1\rangle$,$|r\rangle$ transition, whereas the decay is driven on the $|e_2\rangle$,$|g\rangle$ transition.  \label{fig:pumping_scheme}} 
\end{center}
\end{figure}

In order to illustrate the physical context of the present paper let us 
consider fermionic atoms with four internal states $|g\rangle,|r\rangle,|e_1\rangle,|e_2\rangle$ in state selective, optical lattice traps
as shown in Fig. \ref{fig:pumping_scheme}.
 The lattice for atoms in states $|g\rangle, |e_1\rangle,|e_2\rangle$ is given by an optical standing wave and has
 lattice constant $\frac{\lambda}{4}$. The ground state lattice is shifted by half a lattice constant compared to the
 other two. Atoms in internal state $|r\rangle$ feel a very shallow potential and are delocalized compared to the tightly
 confined atoms in other internal states.  The transition $|r\rangle \leftrightarrow |e_1\rangle$ is driven by a 
laser field with Rabi frequency $\Omega_1$, whereas $|g\rangle \leftrightarrow |e_2\rangle$ is coupled
to a laser with Rabi frequency $\Omega_2$. Spontaneous decay occurs from the two excited levels into the metastable ground states $|r\rangle, |g\rangle$. The optical lattices are deep so that we can assume that only adjacent Wannier wavefunctions $\phi_i^\mu(x), \mu \in e_1,e_2$ and $\phi_i^g$ or $\phi_{i+1}^g$ are optically coupled by the laser fields.
Using this setup non trivial pump and loss processes into and from the metastable states $|g\rangle$ are realized. Atoms in the shallow $|r\rangle$ potential act as a reservoir for the optical transitions. 

First of all atoms from the reservoir $|r\rangle$ are pumped via $|e_1\rangle$  into a superposition of atoms in neighboring lattice sites with fixed relative phase. To see this we note that the spontaneous emission from $|e_1\rangle$ to $|g\rangle$ can be described by the interaction Hamiltonian
\begin{equation}
\Ham_{int}=\int dx \sum_{\vec{k}}\left(g_{\vec{k}}\hat{a}_{\vec{k}}\hat{\Psi}_g^\dagger(x)\hat{\Psi}_{e_1}(x) e^{ik_xx}+h.a.\right),
\end{equation}
where $\hat a_{\vec k}$ is the annihilation operator of the electromagnetic mode with wavevector $\vec k$, and
$g_{\vec k}$ is the corresponding coupling matrix element. 
Using the decomposition of the fermionic fields in internal states $|g\rangle$ and $|e_1\rangle$ into the Wannier basis $\hat{\Psi}_g(x)=\sum_j\phi_j^g(x)\hat{c}_j$ and $\hat{\Psi}_{e_1}(x)=\sum_j\phi_j^{e_1}(x)\hat{e}_{1,j}^\dagger$ yields
\begin{equation}
\Ham_{int}=\sum_j\sum_{\vec{k}}\left[g_{\vec{k}}\hat{a}_{\vec{k}}\hat{e}_j\Bigl(\eta_{j1}^{(k)}\hat{c}_j^\dagger+\eta_{j2}^{(k)}\hat{c}_{j+1}^\dagger\Bigr)+h.a.\right],
\label{eq:Ham-pump}
\end{equation}
where $\eta_{j1}^{(k)}=\int dx\,  \phi_j^{e_1}(x)\phi_j^g(x)e^{ik_xx}$ and $\eta_{j2}^{(k)}=\int dx\,  \phi_j^{e_1}(x) \phi_{j+1}^g(x)e^{ik_xx}$ denote the Frank-Condon factors corresponding to the transitions $j\rightarrow j, j+1$.
Due to the exponentially decreasing Frank-Condon overlaps all transitions with $j'\neq j,j+1$ can safely be neglected. 
As the Wannier functions in a deep optical lattice are well localized the
products $\phi_j^{e_1}(x)\phi_j^g(x)$ and $\phi_j^{e_1}(x)\phi_{j+1}^g(x)$ are well localized functions at positions
$x_j \pm a/4$ with $a \sim \lambda/2$ being the lattice constant.
Thus (\ref{eq:Ham-pump}) can be rewritten as
\begin{equation}
\Ham_{int}=\sum_j\sum_{\vec{k}}\left[g_{\vec{k}}\hat{a}_{\vec{k}}\hat{e}_j\eta_{j1}^{(k)} \Bigl(\hat{c}_j^\dagger+\nu\hat{c}_{j+1}^\dagger e^{i k_x a/2}\Bigr)+h.a.\right],
\label{eq:Ham-pump2}
\end{equation}
where the (real) parameter $\nu$ can be tuned by shifting the position of the $|e_1\rangle$ lattice, relative to that of $|g \rangle$. 
Coupling of the many motional states in $|r\rangle$ with a laser to $|e_1\rangle$ leads after elimination of the vacuum modes to an optical pumping that can be described by independent Lindblad generators
\begin{equation}
 L_j^{pump}=\chi (c_j^\dagger+\nu c_{j+1}^\dagger)
\end{equation}
with $\chi\sim\frac{\Omega_1^2}{\gamma}$, which describe the coupling to two adjacent lattice sites.
Note that the relative phase term $e^{ik_x a/2}$ in Eq.(\ref{eq:Ham-pump2}) vanishes after averaging
over the vacuum modes up to the first order in $k_0 a/2$.

We now show that optical pumping from $|g\rangle$ via $|e_2\rangle$ leads to a loss of fermions in all superpositions of neighboring sites except for one dark mode. The corresponding Hamiltonian describing the coherent part of the interaction reads
\begin{equation}
\Ham_{int,2}=\sum_j\Omega_2 \hat{f}_j\left(\tilde{\eta}_{j1}\hat{c}_j+\tilde{\eta}_{j2}\hat{c}_{j+1}\right)+h.a. 
\end{equation}
where  $\tilde{\eta}_{j1}^{(k)}=\int dx \, \phi_j^{e_1}(x)\phi_j^g(x)e^{iq_xx}$ and $\tilde{\eta}_{j2}^{(k)}=\int dx \, \phi_j^{e_1}(x)\phi_{j+1}^g(x)e^{iq_xx}$ where $q_x=\vec{q}\cdot\vec{e}_x$ and $\vec{q}$ is the wave vector of the laser corresponding to $\Omega_2$. Note that since the wavevector of $\Omega_2$ is well defined $\tilde{\eta}_{j1}$ and $\tilde{\eta}_{j2}$ differ in both amplitude and phase. Considering a fast subsequent decay from $|e_2\rangle$ finally gives rise to an optical pumping out of state $|g\rangle$ described by independent Lindblad generators 
\begin{equation} 
 L_j^{decay}=\gamma\left(c_j+\nu e^{ig} c_{j+1}\right).
\end{equation}
Without loss of generality we can set $\gamma=1$ which fixes the overall time scale of the process. Then the free
parameters of the Liouvillian are the amplitudes $\chi$ and $\nu$ and the phase $g$. 

\begin{figure}
\begin{center}
\includegraphics[width=0.98\columnwidth]{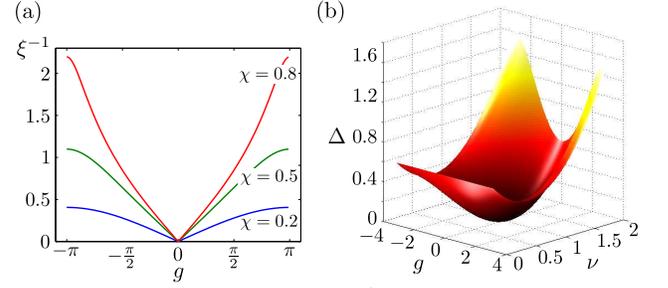}
\caption{(Color online) For the quantum-optical example the inverse correlation length shown in (a) shows a linear behavior around a critical point at $g_c=0,\nu_c=1$. Near the critical point the damping gap $\Delta$, shown in (b), approaches zero with exponent $2$. Both length and time scales diverge in the vicinity of the critical point.
\label{fig:example}} 
\end{center}
\end{figure}

Solving the Lyapunov-Sylvester equation (\ref{eq:Lyapunov}) we find for the symbol function
of the correlation matrix
\begin{equation}
 \gamma(\phi) =\frac{1}{d(\phi)} \begin{pmatrix}
n_{11}(\phi) & n_{12}(\phi) \\
-n_{12}(\phi) & n_{11}(\phi)
                       \end{pmatrix},
\end{equation}
where 
\begin{align}
 n_{11}(\phi) &=4 i \nu  \chi ^2 \left(1+\nu ^2+2 \nu  \cos\phi\right) \sin g \sin\phi,&\\
 n_{12}(\phi) &=\left(1+\nu ^2+2 \nu  \cos g \cos\phi\right)^2&\\
&\!-4 \nu ^2 \sin^2 g \sin^2\phi-\chi^4\left(1+\nu ^2+2 \nu  \cos\phi\right)^2\nonumber
\end{align}
and
\begin{multline}
 d(\phi)=\left[\left(1+\nu ^2\right) \left(1+\chi ^2\right)+2 \nu  \left(\chi ^2+\cos g\right) \cos\phi\right]^2\\
-4 \nu ^2 \sin^2g \sin^2\phi.
\end{multline}
The Fourier-transform according to Eq.(\ref{eq:correlation}) 
yields the correlations of Majorana fermions. From the symmetry of
$\gamma(\phi)$ it is immediately clear that as expected only normal correlations of
fermionic creation and annihilation operators are non zero.
Following Ref.\cite{Eisert2010b} we can define criticality by a diverging correlation length
\begin{equation}
\xi^{-1}  =-\lim_{|i-j|\rightarrow \infty}\frac{\log|\langle \hat c_i^\dagger \hat c_{j}\rangle_{\rm SS}|}{|i-j|}.
\end{equation}
The correlations $\langle\hat c_i^\dagger \hat c_j\rangle$ become infinitely long ranged for $\nu=\nu_c=1$ and $g=g_c=0$ and any non-vanishing value of $\chi$.
In Fig. \ref{fig:example}(a) the dependence of $\xi$ on the phase $g-g_c$ is shown for $\nu=1$. 
One recognizes a linear
dependence and the same holds for the dependence on $\nu-\nu_c$. Hence the critical exponent for this
example is $\lambda=1$. The same critical exponent has been found by Eisert and Prosen in \cite{Eisert2010b}.

The NESS with correlation matrix $\Gamma_0$ is an attractor of the dynamics, i.e. small deviations 
$\delta \Gamma$ from it will decay to zero after some time.  An important quantity is the smallest decay rate for such deviations
as it defines
the time-scale of decay back to the stationary state. Since close to the critical point correlations become infinitely long ranged
one expects that the time scale for re-establishing long-range order after a small perturbation tends to infinity. This corresponds
to a closure of the damping gap $\Delta$, i.e. the smallest non-vanishing eigenvalue of the real part of the Liouvillian. The damping gap
$\Delta$ is here the direct counterpart to the excitation gap $\Delta_H$ in unitary systems.
Staying in the manifold of Gaussian states, the dynamical equation for $\delta\Gamma$ reads
\begin{equation}
 \frac{d}{dt}\delta\Gamma = X^T \,\delta\Gamma +\delta \Gamma \, X.
\end{equation}

Thus one has to consider the eigenvalues of $X$, i.e. $R+R^*$. One again finds that
these eigenvalues vanish for arbitrary values of $\chi$ for $\nu=\nu_c=1$ and $g=g_c=0$.
Fig. \ref{fig:example}(b) shows the damping gap for the present example. One recognizes a closure
at the isolated point  $s_c=\nu_c e^{i g_c} =1$ with a quadratic dependence in the immediate
vicinity. Thus the dynamical critical exponent is $z=2$. 

In this particular example we have seen that a critical point is associated with both a vanishing of the damping gap $\Delta$ 
and a diverging correlation length. 
One recognizes furthermore that although the parameter space of the reservoir coupling is two-dimensional (disregarding
the irrelevant parameter $\chi$), there is only a single point where the system is critical. The parameter
region with nonvanishing $\Delta$ is a connected manifold
and thus there is only one distinct gapped phase. Continuous changes of the parameter $s=\nu e^{ig}$ in the complex plane circumventing
the critical point $s_c=1$ will smoothly connect the entire gapped region.

The present example gives rise to a number of questions: Is a diverging correlation length always connected
to a vanishing damping gap? What are the possible universality classes, characterized by the
critical exponents $\lambda$ and the dynamical critical exponents $z$? In the following we will discuss these issues for linear fermion models with general local reservoir couplings.


\section{free lattice fermions with general local reservoirs}

\begin{figure}
\begin{center}
\includegraphics[width=0.90\columnwidth]{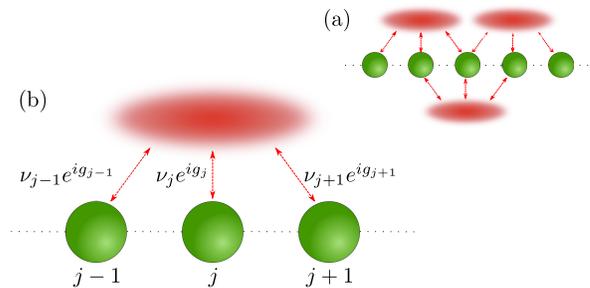}
\caption{(Color online) (a) Schematic representation of a linear chain translationally invariant coupled to independent reservoirs. (b) The single reservoir is characterized by $2N$ real parameters as indicated for a three site reservoir coupling.  \label{fig:scheme}} 
\end{center}
\end{figure}

Let us now consider fermionic lattice models with reservoirs
that couple simultaneously to $N$ adjacent lattice sites.
A schematic representation of our model is shown in Fig. \ref{fig:scheme}.
To simplify the calculations and the final expressions we restrict ourselves
to Lindblad generators which only contain a single type of Majorana fermions. 
A generalization is however straightforward
and all conclusions hold. In particular we consider only Majorana operators of the first kind ($w_{2j-1}$) in the generators as they are invariant under the exchange of creation and annihilation 
operators. 
The dissipative dynamics is decoupled from the even Majorana modes, leading to degeneracies in the NESS. This degeneracy can be lifted by a free, translation invariant Hamiltonian and the addition of this unitary term does not change the properties of the uneven Majorana modes. In this case the steady state equation (\ref{eqn:lindblad_steadyI}) has a trivial solution in terms of the symbol function $r(\phi)$, which represents the reservoir coupling
\begin{align}
r(\phi)&=\sum_{m,n}\nu_m\nu_ne^{-i\phi(m-n)} e^{i(g_m-g_n)},\\
 \gamma(\phi)&=\frac{r(\phi)-r(-\phi)}{r(\phi)+r(-\phi)}\mathbbm{1}_{2\times 2}. \label{eqn:symbolf}
\end{align}
Apart from the fact that the Hamiltonian guarantees its uniqueness, the NESS
is independent of the Hamiltonian details and therefore possible critical features of the ground state of ${\cal H}$ 
are not recovered in the steady state. Moreover (\ref{eqn:symbolf}) does in general not correspond to a pure state. The particle-hole symmetry of the Lindblad generators 
leads to a mean occupation of $\frac{1}{2}$ in the NESS.
The eigenvalues of the circulant correlation matrix $\Gamma$ are given 
by the entries of the symbol function $\gamma(\phi)$ which is positive. They are bounded between $[-1,1]$, which 
is in contrast to pure states for which it can be shown that all eigenvalues must be $\pm 1$.

 By inverse Fourier transform we calculate real-space correlations in the steady state 
\begin{equation}
\langle 
w_1w_{1+2d}\rangle=\frac{1}{2\pi}\int_0^{2\pi}\! \! d\phi\ e^{i\phi d}\gamma_{11}(\phi).
\end{equation}
The integration can not be carried out in general but we can understand the characteristic properties for large spatial distances $d$ by using arguments of complex
calculus. To this end we rewrite the symbol $\gamma(\phi)$ as a function in the complex plane using the substitution $e^{i\phi}=z$
\begin{align}
\gamma_{11}(z)&=i\frac{\sum_{j,l}\nu_j\nu_le^{i(g_j-g_l)}(z^{j-l}-z^{l-j})}{\sum_{j,l}\nu_j\nu_le^{i(g_j-g_l)}(z^{j-l}+z^{l-j})}\label{eqn:symbolf_z},\\
 \langle w_1w_{1+2d}\rangle&=\sum_{a\in S_1}\text{Res}_a\left[z^{d-1}\gamma_{11}(z)\right],\label{eqn:residue}
\end{align}
where Res$_ a$ denotes the residues inside the unit circle $S_1$, ($|z|\le 1$).
The residue is non-zero only in singular points of $\gamma(z)$. Because numerator $n(z)$ and denominator $d(z)$ of (\ref{eqn:residue}) are holomorphic, only zeros of the denominator inside the unit circle contribute to the correlations. 
The symbol function has at most simple poles and it has been pointed out in \cite{Eisert2010b}, that the zero closest to the unit circle is relevant for the large $d$ behavior. As the denominator $d(z)$ of Eq.(\ref{eqn:symbolf}) is just the symbol function of the matrix $X$, zeros of $d(z)$ on the unit circle correspond to both a diverging correlation
length {\it and} a closure of the damping gap $\Delta$, making both definitions of noise-induced criticality for
linear models identical. 

\subsection{conditions for criticality}

The critical points, i.e. the singularities of the symbol function are the roots of $d(z)$ on the unit circle.
For general reservoir couplings to multiple sites, explicit expressions for the roots of $d(z)$ are hard to 
obtain and we need a different criterion to find the critical parameters.
On the unit circle the denominator $d(z)$ is strictly non-negative because it can be rewritten as $
d\big(z,\{g_j\},\{\nu_j\}\big)=|\sum_j\nu_j e^{ig_j}z^{j}|^2+|\sum_j\nu_j e^{ig_j}z^{-j}|^2$. A configuration 
$\{s_j\}= \{\nu_j e^{ig_j}\}$ of the complex parameters of the Liouvillian leads to a critical behavior if a $z_0$ on the unit circle exists, such that 
the individual sums inside the absolute value vanish for $z_0$ and its complex conjugate. This gives a pair of implicit equations 
\begin{equation}
1+\sum_ {j=1}^{N-1} s_j\, z_0^j = 1+\sum_{j=1}^{N-1} s_j \, z_0^{j*} =0, \label{eqn:critical-point}
\end{equation}
where we have used that without loss of generality $s_0$ can be set equal to unity.
Apparently reservoir couplings to a single site $(N=1)$ cannot induce criticality, however couplings to
$N>1$ sites may. For a given $z_0$, $2(N-2)$ out of the $2(N-1)$ real parameters $\nu_j, g_j$  can be chosen
arbitrarily. As there are only a finite number of roots $z_0$ 
the non-trivial complex solutions $\{s_j = \nu_j e^{ig_j}\}$ to these equations are limited to a $d-2$ dimensional
manifold in the $d=2(N-2)$ dimensional parameter space. As a consequence there can never be two extended,
non connected regions in parameter space with a finite damping gap $\Delta$. Thus linear systems can become critical,
but there are no reservoir-induced phase transitions between distinct gapped phases.


\subsection{correlation length and critical exponents}


In the vicinity of the critical point the behavior of $\xi$ is determined by the leading-order exponent $\lambda$ of the singularity, which itself is determined by the properties of the denominator $d(z)$ in Eq.(\ref{eqn:symbolf_z}). If a zero of $d(z)$ approaches the unit circle from the inside, $\xi$ diverges which corresponds to a phase transition to criticality.
Let $z_0$ be the closest singularity to the unit circle, then the correlation length is given by 
\begin{equation}
\xi^{-1}=-\ln|z_0|\approx 1-|z_0|.
\end{equation}
In the following we will analyze the dependence of $\xi$ on the system parameters in the vicinity
of such singularities and determine the corresponding critical exponents.

For two-site coupling the implicit equations can have a nontrivial solution and we find that the generator 
$L_{\text{two}}=\nu_0e^{ig_0}w_1+\nu_1e^{ig_1}w_3$ leads to a critical NESS 
for $g_0-g_1\in\pi\mathbbm{Z}$ and $\nu_0=\pm\nu_1$. Analyzing the behavior in the vicinity of the critical points 
we find a critical exponent of $\lambda=1$ in agreement with the results of \cite{Eisert2010b}. The question arises if this value is the only possible one in free fermionic lattice models. To answer this question let us
consider a more general reservoir with coupling to three adjacent sites: $L_{\text{three}}=L_{\text{two}}+\nu_2e^{ig_2}w_5$. 

A possible solution of the implicit equations (\ref{eqn:critical-point}) is $\nu_0=\nu_1=\nu_2$ and 
$g_0=-\frac{2\pi}{3}, g_1=0,  g_2=\frac{2\pi}{3}$. Under variation of for example $g_1$ the critical exponent is
here again $\lambda=1$. However, there is another solution
$g_0=g_1=g_2$ and $\nu_0=\frac{1}{2}\nu_1=\nu_2$. In this case the critical exponent under variation of
$g_1$  is different and given by $\lambda=\frac{1}{2}$.

\begin{figure}
\begin{center}
\includegraphics[width=0.98\columnwidth]{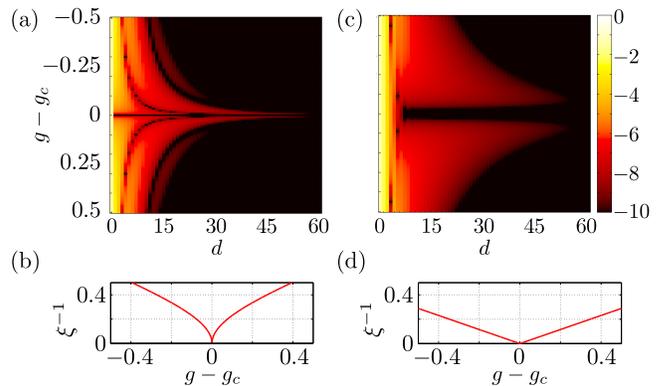}
\caption{(Color online) (a) and (c) show two site correlations $|w_1w_{1+2d}|$ on a logarithmic color scale in dependence of the spatial distance $d$ and a parameter change around critical phase $g_c$. The corresponding inverse correlation length $\xi^{-1}$ is plotted in (b) and (d) and reveals the critical exponent. The left side is the $\lambda=\frac{1}{2}$ transition and the right side the $\lambda=1$ transition for the chain with three site reservoir coupling discussed in the text.
\label{fig:threeSiteColor}} 
\end{center}
\end{figure}

In Fig. \ref{fig:threeSiteColor} we have plotted the two-point correlation functions for sites separated by a distance $d$ in dependence of the phase $g$ that drives the transition for the case of a three-site reservoir coupling. The left part of the figure corresponds to the $\lambda=\frac{1}{2}$ case, whereas the right part corresponds to a transition with $\lambda=1$. The analytic behavior of the inverse correlation length around the critical points, shown in the lower parts of the plots, clearly distinguishes the two cases. 

One recognizes that the amplitude of the component with long-range correlations vanishes as one approaches the critical points. More precisely
for the case of the smaller critical exponent, $\lambda=1/2$, all non-local correlations vanish, while for the case of the larger critical exponent, $\lambda=1$, some component with finite correlation length remains also at the
critical point. 

In the following we want to get some general insight into what are the possible critical exponents 
of a reservoir-driven phase transition. 
As we are interested in the behavior in the vicinity of the critical point, we need to expand the denominator 
$d(z)$ of the symbol function in terms of the relevant system parameter $\{g_j\},\{\nu_j\}$ around their 
critical values. Since we do not have an explicit expression for the roots $z_0$ of $d(z)$, we need to do this in an 
implicit way, i.e. expanding $d(z)$ both in terms of $z-z_0$ and in their explicit dependence on $\{g_j\},\{\nu_j\}$.

Due to the positivity of $d(z)$ on the unit circle, all first order partial derivatives with respect to 
$z$ as well as to the system parameter $g_k$ and $\nu_k$ must be zero at the critical point. 
To find the leading order expansion in $z-z_0$ we thus evaluate higher order partial derivatives with respect to $z$ using the implicit equations (\ref{eqn:critical-point})

\begin{eqnarray}
 \frac{\partial^2 d}{\partial{z^2}}&=&z^{-2}\sum_{j,l}\nu_j\nu_l e^{i(g_j-g_l)}\Big[(j-l)(j-l-1)z^{j-l}\nonumber\\
&&+(l-j)(l-j-1)z^{l-j}\Big].\label{eq:second-derivative}
\end{eqnarray}
If the second order derivative is nonzero at $z=z_0$, we can stop at this level. On the other hand, the second order derivative 
vanishes if a second pair of independent implicit equations is fulfilled, which can easily be read off from (\ref{eq:second-derivative}). This procedure can be continued and each term $\frac{\partial^{2m} d}{\partial z^{2m}}\Big|_{z_0}$, which is zero, yields a new pair of implicit equations

\begin{align}
 \sum_j\nu_jj^m e^{ig_j}z_0^j=\sum_j\nu_jj^m e^{ig_j}z_0^{*j}=0. \label{eqn:implicitm}
\end{align}
Here we have used that the first non-vanishing derivative must be an even one.
Let us assume that all derivatives in $z$ vanish up to order $2M-1$. Then it can be shown, that mixed
derivatives of the type $\partial_{(\nu_k,g_k)}\partial_z^m d\Big|_{z_0}$ vanish for all $m<M$. Thus what remains 
are the second order partial derivatives with respect to $\nu_k$ or $g_k$. Second order partial derivatives in the same parameter are always non zero on the unit circle
(except for trivial cases), as $\partial_{\nu_k}^2 d(z)\Big\vert_{z_0} = 2,\quad  \partial_{g_k}^2 d(z)\Big\vert_{z_0}
=2\nu_k^2$. Thus we can write the power expansion of $d(z)$ in the following general way

\begin{equation}
\begin{split}
 d(\tilde{z},\tilde{x})\approx C_2 \tilde{x}^2+\tilde{x}
\Big[C_{1,\tilde{M}}\tilde{z}^{\tilde{M}}+\mathcal{O}(\tilde{z}^{\tilde{M}+1})\Big]\\
+C_{0,2\tilde{M}}\tilde{z}^{2\tilde{M}}+\mathcal{O}(\tilde{z}^{2\tilde{M}+1}),
\end{split}
\end{equation}
where $C_2$ and $C_{0,2\tilde{M}}$ are non zero constants. Here
$\tilde{x}$ is a linear combination of parameter variations from the critical values $\tilde{g}_k=g_k-g_{kc}$ and $\tilde{\nu}_k=\nu_k-\nu_{kc}$,  and $\tilde{z} = z-z_0$. The lowest non-vanishing contributions determine the critical exponent and therefore $\tilde{M}$ is the minimal $M$ of all parameters included in $\tilde{x}$.
The zeros $z_0$ are algebraic functions of the system parameters and we can therefore write $\tilde{z}\approx \tilde{x}^{\lambda}+\mathcal{O}(\tilde{x}^{\lambda+1})$. At least two terms in the expansion must be of the lowest order and therefore we find $\lambda=\tilde{M}^{-1}$ along the line $\tilde{x}$ if the first $\tilde{M}$ implicit equations are fulfilled. We see that all possible critical exponents are the inverse of integer numbers.
The smallest possible critical exponent is determined by the maximum $\tilde{M}$, which is just given by $M$.

We now can relate the $M$ to the number $N$ of adjacent sites coupled by each local reservoir. It is clear that equations ($\ref{eqn:implicitm}$) are linearly independent for different $m$. On the other hand only $N$ equations can be independent for a finite reservoir coupling. This proves that $M$ can be at most $N-1$ or all orders vanish, in which case the symbol function must be zero everywhere and the system is not critical.
We conclude that if the reservoir coupling is restricted to $N$ sites, the critical exponent is out of a bounded set of fractional numbers
\begin{equation}
\lambda\in\left\{1,\frac{1}{2},\frac{1}{3},\cdots, \frac{1}{N-1}\right\}.
\end{equation} 
This is the main result of the present paper.

The corresponding Taylor expansion of the numerator $n(z)$ in a critical point is of higher order than the denominator. Therefore the amplitude of the critical correlations, vanishes as $\xi$ diverges. This is seen in the graphs of Fig. \ref{fig:threeSiteColor}. The remaining non-local correlations, visible for example
in part b) of that figure are due to additional singularities inside the unit circle. Only in the case of the minimal critical exponent $\lambda=\frac{1}{N-1}$, these singularities cannot exist due to fundamental laws of algebra. In
this case the NESS is completely mixed in the critical point.

Another result that can be drawn from our analysis is the dimensionality of the critical parameter space. 
Critical points have to fulfil the set of equations (\ref{eqn:implicitm}) and for a given critical exponent the corresponding dimensionality is given by 
\begin{equation}
 \text{dim}(P_{\lambda})=2\left(N-1-\frac{1}{\lambda}\right).
\end{equation}
It is clear that the critical points are always a zero measure subset of parameter space, but they are not necessarily isolated points, with the exception of critical points with minimal exponent for the given configuration, which are always singular.

\subsection{spectral gap of relaxation rates and dynamical critical exponent}

For unitary lattice models it is well established that the presence of a finite gap in
the excitation spectrum leads to a finite correlation length, while the transition to
criticality is associated with a vanishing gap \cite{Hastings2006}.
In the following we want to establish a corresponding relation for reservoir driven
phase transitions and discuss the dynamical critical exponents.

The relaxation rates of the system are determined by the homogeneous part of (\ref{eqn:lindblad_cov}) and therefore the damping matrix $X$. More precisely the damping matrix $X$ describes the dynamics on the submanifold of Gaussian states, whereas the full system is spanned by the Liouville Operator $\mathcal{L}$ in equation (\ref{eqn:lindblad_master}). The trace preservation of the Lindblad dynamics requires $\mathcal{L}$ to have at least one eigenvalue with vanishing real part. If this zero eigenvalue is unique, the gap in the real spectrum sets the slowest relaxation rate for arbitrary initial states. The gap $\Delta$ of $X$ thus gives an upper bound for the gap of $\mathcal{L}$, and the gap in the full damping spectrum must vanish as one approaches the critical points.
The eigenvalues of $X$ are purely real, when neglecting the Hamiltonian contributions and strictly negative. In the translation invariant system the eigenvalues are given by the symbol function $-r(\phi)-r(-\phi)$, which we have identified before as relevant for the correlation length. In the vicinity of a critical point, the slowest relaxation rate, defining the spectral gap of relaxation, is determined by the roots $z_j$ of $d(z,\{g_j\},\{\nu_j\})$ closest to the unit circle
\begin{equation}
 \Delta=\text{min}_{|z|=1} d(z,\{g_j\},\{\nu_j\}).
\end{equation}
Therefore the dynamical exponent is immediately related to the critical exponent $\lambda$. The exponent of the divergence however must be modified by the number of roots $\kappa_c$, that merge at the same point on the 
unit circle when the Liouville parameters approach their critical values:
\begin{equation}
 \Delta\sim |g-g_c|^{\kappa_c\lambda}.
\end{equation}
The number of roots $\kappa_c$ is thus identical to the dynamical critical exponent $z$. 
For the minimum critical exponent $\lambda=1/(N-1)$ all $2(N-1)$
complex roots merge simultaneously on the same point and thus
\begin{equation}
 z\, \lambda =2,\qquad{\textrm{for}}\quad \lambda = \lambda_{\rm min} \label{eq:z}
\end{equation}
Moreover for all examples we have considered we found that the
damping gap closes as a quadratic function, which suggests that (\ref{eq:z})
is more general.

\section{summary}

To summarize we have analyzed the non equilibrium steady state of translation invariant chains of free fermions coupled to
local Markovian reservoirs described by linear Lindblad generators. Such couplings can be generated e.g. in ultra-cold
atomic lattice gases as we have shown for the example of the two site coupling. 
A general expression for the correlation matrix of the NESS can be obtained using a symbol function ansatz. We showed that under certain conditions the NESS
goes into a critical state upon changing reservoir parameters. The critical state is characterized by a simultaneous divergence of a correlation length and a critical slow-down of relaxation i.e. a closing of the gap in the damping spectrum. 
We showed that the dimension of the critical parameter space is at most $d-2$, where $d$ is the dimension
of the full space of reservoir parameters. As a consequence all gapped phases are smoothly connected and although there
is a transition into a critical phase there is no reservoir-induced quantum phase transition between distinct gapped phases for linear models.
The transitions to a critical state can be classified by the leading order exponent in the dependence of the inverse correlation length on the system parameter that drives the transition. We have shown that this critical exponent must be the inverse of an integer between 1 and ${N-1}$, where $N$ is the number of sites coupled by a single reservoir. 
Furthermore a general expression for the dynamical critical exponent, describing the
closure of the damping gap was derived.

\

The authors gratefully acknowledge financial support from the DFG through SFB-TR49.

\bibliography{jabref.bib}
\end{document}